# Conditional Handover in 5G – Principles, Future Use Cases and FR2 Performance

Jędrzej Stańczak[1,3], Umur Karabulut[2] and Ahmad Awada[2]
[1]Nokia Standards, Poland, [2]Nokia Research, Germany, [3]Wroclaw University of Science and Technology, Poland
E-mail: jedrzej.stanczak@nokia.com

*Abstract*—This paper elaborates on Conditional Handover (CHO) - a mobility feature designed in Release 16 of 3rd Generation Partnership Project (3GPP), aimed at improving the reliability of handover in cellular networks. CHO has turned out to be a successful feature and attracted several other areas of mobile networks, where increased reliability is also desirable (e.g. Non-Terrestrial Networks or Integrated Access Backhaul). This paper explains the principles of CHO, describes its future use in vertical areas and provides performance results showing the gains CHO may offer to 5G systems at higher frequency bands. Our simulative analysis was focused on the handover in Frequency Range 2 (FR2) and CHO recovery procedure. Results have shown that also at higher frequencies CHO can substantially reduce the mobility failures compared to baseline handover of Release 15. Moreover, it has been verified that CHO recovery can be a useful means to allow faster reconnection to the network after a failure. In certain scenarios CHO Recovery usage ratio was found to exceed 80%. The paper also outlines the future research and standardization directions in the area of CHO which appears to be a solid candidate for further development in 3GPP Release 18 and beyond.

## I. INTRODUCTION

Conditional Handover (CHO) has been designed as a part of 3GPP Release 16, finalized in 2020. The main motivation behind standardizing such functionality was to improve the reliability of handover - an essential procedure in cellular networks which shall offer robust and seamless mobility to the users. CHO ensures that by preparing the handover in advance. The early provided configuration is actually taken into use by the User Equipment (UE) only when the associated condition is fulfilled [1]. Thanks to decoupling the preparation and execution phases, the network may safely reach the UE when the user is still experiencing favourable radio conditions in the source cell and the UE is instructed to execute the handover when the candidate target cell becomes good enough in terms of the radio signal quality. As such, the likelihood of experiencing a failure in the source cell or during Random Access (RA) attempt to the target cell is reduced which provides mobility robustness for the UE.

Increased handover reliability has made this functionality attractive to other vertical features developed as a part of 3GPP Release 17 and onwards. For example, the work towards enabling the cellular support via Non-Terrestrial Networks (NTN), initiated in the third quarter of 2020, addresses mobility challenges as one of its primary goals [2]. CHO, as standardized in Release 16, with certain NTN-specific enhancements, is considered to resolve the NTN mobility issues. Another area, where CHO is regarded as a promising mobility technique is Integrated Access Backhaul (IAB), integrating wireless backhaul links with radio access via New Radio (NR). A need to handover an IAB Mobile Termination (IAB-MT) in a reliable and predictable way, creates relevant circumstances to adopt CHO also to this area of cellular communications. Yet another 3GPP feature, where the use of CHO is considered, is NR in unlicensed spectrum (so-called NR-U). NR-U allows the 3GPP radio to operate in the shared spectrum, with the Listen Before Talk (LBT) and other mechanisms known from the technologies where the bandwidth is not reserved for UE's exclusive usage. Due to the need to assess the channel occupancy, it may be beneficial to employ CHO as the number and duration of steps to be performed by the UE before executing the handover can be minimized, which reduces mobility failures.

The paper is organized as follows. Section II provides a comprehensive description of CHO defined as a part of 3GPP Release 16. In Section III the potential use of CHO in other vertical areas, such as NTN or IAB, is studied. Performance results in Frequency Range (FR) 2, including CHO recovery, are presented and discussed in Section IV. Finally, Section V echoes the main findings and elaborates on the next steps.

## II. CONDITIONAL HANDOVER IN 3GPP RELEASE 16

### A. Basics of Conditional Handover

CHO, as the name implies, is executed only if the associated condition is met. That is a principal difference in comparison to the baseline handover (defined in 3GPP Release 15) which is performed immediately upon the reception of handover command. For CHO the UE is configured with up to two execution conditions per each candidate target cell. In case there are two conditions provided, those are linked with conjunction relationship, so both need to be fulfilled before the handover execution may be initiated. 3GPP standard restricts these two execution conditions need to be configured for the same Reference Signal (RS), but can use different measurement quantities for evaluation, such as Received Signal Reference Power (RSRP), Received Signal Reference Quality (RSRQ) or Signal to Interference and Noise Ratio (SINR). Execution conditions in the form of measurement Event A3 (neighbour cell becomes offset better than serving cell) or Event A5 (serving cell becomes worse than 1st threshold and neighbour cell becomes better than 2nd threshold) [6], are the same as used for Measurement Report (MR) triggering. For CHO, however, there will be no reporting to the serving cell, instead a handover is executed towards the target cell.

Fig. 1 depicts the basic steps of Conditional Handover, defined by 3GPP. In Step 1 the UE sends to the source 5G base station (Source gNB in Fig. 1) a Measurement Report comprising the results for potential candidate target cells.

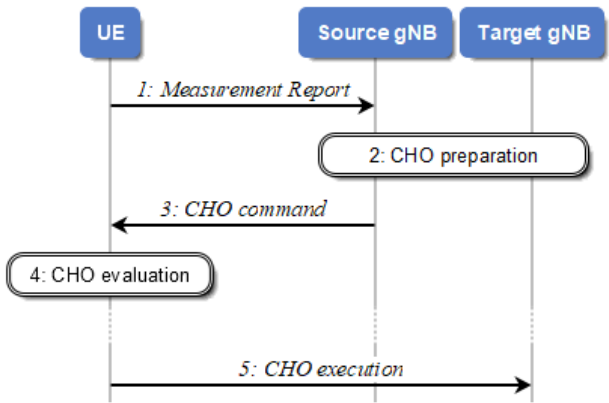

Fig. 1: Basic steps of Conditional Handover

Source gNB asks selected Target gNBs to acknowledge the handover request and prepare corresponding configurations, to be used by the UE for accessing the target cell (Step 2). There may be up to eight candidate target cells prepared, whereas for simplicity Fig. 1 shows just a single Target gNB. In Step 3 the Source gNB sends a CHO command (Radio Resource Control (RRC) Reconfiguration [6]) comprising the configurations and execution conditions for the cells that have been acknowledged in Step 2. There is a clear split of roles defined in CHO: source gNB is responsible for setting the CHO execution conditions, while the target gNB provides the CHO command containing the target cell-specific configuration. Upon reception of CHO command (Step 3) the UE does not detach from the Source gNB - unlike in the baseline handover. Instead, the UE continues exchanging the Uplink (UL) and Downlink (DL) data with the Source gNB until the CHO execution condition is met. This occurs during Step 4, shown in Fig. 1. Once the execution condition for any configured and prepared cell is fulfilled, the UE moves to Step 5 and initiates CHO execution. Herein, the UE follows the same procedure as it would apply for the baseline handover, i.e. sends the RA preamble and awaits the Target gNB's response. Timer T304, controlling the execution of handover [6], is started only at Step 5, while the timer T310, controlling the radio link in the serving cell, is stopped at Step 5. In the rest of the sequel, *CHO preparation* and *CHO execution* corresponds to triggering Fig. 1's Step 1 and 5, respectively.

*B. Basics of Conditional Handover Recovery*

Although the UE may be prepared with multiple candidate cells, the UE attempts to access just a single target cell, the one that was first to fulfil the execution condition. If more than a single cell meets the execution condition, it is up to the UE to decide which cell to access. The UE is not allowed to try the handover execution multiple times or to execute several RA attempts in parallel, towards different candidate cells. Despite providing the UE with CHO commands early and configuring it to execute the CHO late (terms "early" and "late" are relative to baseline handover), it may occur the CHO is not successful. In such circumstances, according to [6], handover failure is declared and the UE performs a connection reestablishment procedure. One can easily notice it would be suboptimal to follow the legacy reestablishment process, resulting in a long data interruption, if the UE was prepared with multiple candidate cells and the access to just one of them has failed. Thus, 3GPP Release 16 has specified a mechanism allowing UE to make use of those stored configurations to recover from that failure. The UE performs cell selection, which is an inherent part of reestablishment procedure.

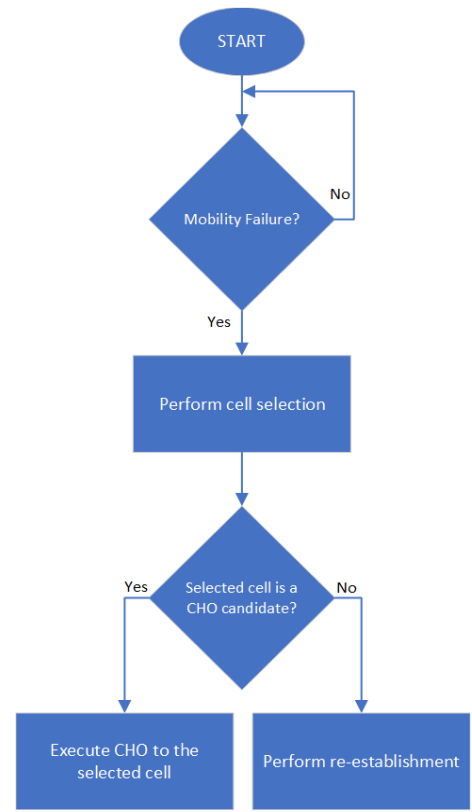

Fig. 2: Link recovery using Conditional Handover configurations

If the selected cell is a CHO candidate cell, the UE may access it by applying the stored configuration and performing handover, abandoning the reestablishment procedure. Otherwise, the UE performs reestablishment. This is illustrated in the block diagram in Fig. 2. By applying this procedure, the UE may accelerate recovery actions and avoid the time-consuming radio link reestablishment process. The performance of this newly defined recovery mechanism using cells that are prepared with CHO is further analysed in Section IV.

III. CHO IN VERTICALS AND OTHER AREAS

As the usefulness of CHO was widely recognized in 3GPP, analysed and proven in various conference papers (e.g. [3][4]), the solution has been considered for other areas, where cellular communication is applied. In this chapter the possible adoption of CHO in vertical domains, i.e. for other use cases than Mobile Broadband (MBB), will be described.

*A. CHO in Non-Terrestrial Networks*

The support of cellular technology using satellites brings new opportunity to offer global coverage also in the areas which were so far deprived of mobile communications. However, this comes at certain expense – at least the need to resolve demanding NTN-specific issues. One of such areas to address is mobility, which in Non-Terrestrial Networks could be problematic, especially for Low-Earth Orbit (LEO) satellites, typically operating at the altitudes between 600 and 1200 kilometres. This alone implies large propagation delays which inevitably affects the on-time execution of the HO. Furthermore, LEO satellite's continuous movement results in both the UEs and the gNBs operating in motion, unlike in Terrestrial Networks (TNs), where the base stations' position is fixed. One can easily conclude this leads to the temporary nature of the coverage provided on Earth, which will move along with satellites movement. LEO satellites' velocity can be roughly 7.5 km/s, which is clearly a non-negligible value in terms of how quickly the footprint on Earth will disappear. The authors of [5] provide a thorough analysis of HO performance in NTN.

To mitigate the aforementioned challenges, CHO has been studied and decided to be specified with NTN-specific enhancements [2]. In the classical approach to CHO, the execution criteria are evaluated in terms of the radio condition (e.g. received signal strength/quality) of the source and target cells. For NTN, however, new conditions would need to be introduced, to consider the satellite's movement and temporary coverage (as shown in [5], HO in LEO-based NTN is attempted every 5-6 seconds). Relying on the received signal strength/quality distribution within the cell may be insufficient as the deviation of measured received signal in the cell centre and cell edge may be too minor for the UE to realize the cell's signal strength/quality is deteriorating and the coverage will disappear soon. Thus, it was decided to introduce two additional handover execution criteria for NTN use case [1]:

- Timing-based handover execution
- Location-based handover execution

As the satellite's motion, including velocity and direction, is predictable, it is possible to determine when particular cell's coverage will be available on the Earth ground. By using such knowledge, the network may configure the UE with CHO execution condition which defines a time window [t1, t2] within which the UE is allowed to attempt accessing the associated cell. Naturally, limiting the entire CHO execution criteria to the condition verifying the timing could be detrimental if the radio conditions are not checked simultaneously. Thus, the most reasonable implementation of time-based CHO triggering is to configure it jointly with radio-based measurement event (e.g. Event A3, A4 or A5, as defined in [6]). The UE evaluates the radio-based execution condition after time t1 and if it is met prior to time t2, UE executes CHO to the selected candidate cell.

A somewhat similar background can be revealed for using the location to trigger the CHO in NTN systems. The UE may be configured to verify the distance to a reference point, associated with each of the NTN cells (i.e. current serving and potential target cell). If a single beam from the satellite is used to provide the cell coverage, then the reference location may be defined as beam centre. The condition for location-based CHO triggering could be defined as an inequality, where the distance between the UE and beam centre for source cell becomes larger than threshold and the distance between the UE and the target cell shall become lower than threshold. Similarly to timing-based CHO execution, using location as the only factor for triggering cell change could be a risky approach, so a desirable practical use is to apply the location-based event jointly with radio measurement based criteria.

In basic approach CHO is used for preparing the imminent handover (i.e. only for a single cell change). It cannot be used for configuring distant cells, to be accessed in next few hops. In NTN, however, the early preparation for the future access may be feasible as the cell sequence could be predicted. Satellite move over orbits in a predefined way, whereas the UEs are relatively static. This may enable the configuration of chain of CHOs, where the UE is prepared in a single procedure with CHO configurations for multiple steps ahead.

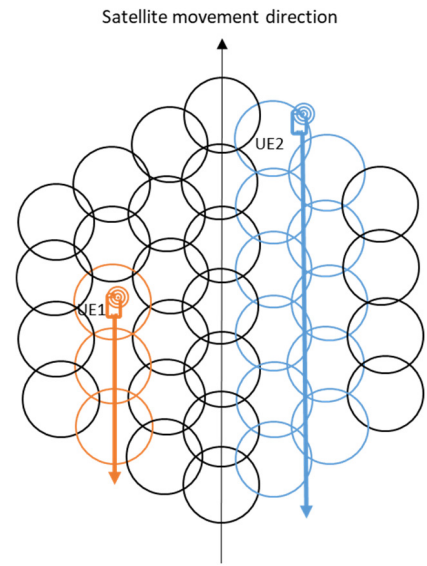

Fig. 3: The example of the deployment where CHO chain may be used

Fig. 3 depicts how such chain of conditional handovers may look like. Orange and blue arrows reflect the UE's and coverage motion trajectory, while circles are used to denote the expected visited cells by UE1 and UE2, respectively. When entering RRC Connected state, the UE may be provided with the CHO configurations for all the cells in the estimated path, as shown in Fig. 3. This allows to save on radio interface signalling and - if combined with smart resource reservation techniques at the network side - excessive radio resource pre-assignment may not be critical. In addition, UEs will benefit from reduced ping-pong rates (i.e. returning to the source cell soon after HO execution) and are less prone to experiencing mobility failures.

*B. CHO in Integrated Access Backhaul*

Integrated Access Backhaul is a functionality in New Radio, which allows to relay the wireless communication using Radio Access Network (RAN). A new entity, called IAB-donor, has been introduced, whose responsibility is to backhaul the traffic and additionally offer the gNB functionality. IAB-donors are connected with IAB-nodes, which are the equivalents of gNB-DU (Distributed Unit) and also support the UE functionalities (denoted in this context as IAB-MT). The detailed architecture and related procedures are captured in the dedicated technical report [7]. 3GPP Release 17 specifies further enhancements to IAB framework, including CHO for improved reliability. CHO is considered for IAB-MTs, which shall largely resemble the procedure defined in Release 16 for handing over the UEs. However, CHO is also a candidate solution for IAB-node migration, where IAB-node moves to a different parent node under the same IAB-donor CU (Centralized Unit). IAB-node is static, so the condition for executing the migration will likely differ from the one defined for IAB-MT. The 3GPP Release 17 work concerning the enhanced IAB will be finalized in the middle of 2022.

*C. CHO in NR-Unlicensed*

Mobility in NR-U is another area which has to be tackled differently than handovers in typical systems operating in the licensed frequency bands. In NR-U multiple UEs may be competing for the access to the spectrum in an uncoordinated manner. Thus, it may be insufficient to reuse the existing Release 16 CHO triggering conditions which rely purely on reference signal quality or power level. Instead, it shall be considered how to employ the LBT result into the decision to execute CHO. The UE shall sense the communication channel and trigger the CHO only if the result of such channel verification meets the predefined criteria. If the structure of the measurement events defined in [6] shall be followed, the proposed entry condition for the corresponding event could be determined as:

$$M_c - Hys > Threshold \quad (1)$$

where $M_c$ is the measurement result for channel occupancy related to candidate CHO cell, *Hys* represents the hysteresis applied to the result, while *Threshold* is the channel occupancy reference level, which shall be exceeded for the inequality to be met. The condition (1) relates just to the channel occupancy and should preferably be associated with other criteria for executing the CHO, such as Event A3, A4 or A5 [6]. The principles for CHO in NR-U described above are not yet a part of the global 3GPP cellular standard. However, these will be taken into account when mobility robustness in unlicensed spectrum is enhanced.

## IV. PERFORMANCE RESULTS

This section presents the performance of CHO in FR2 and CHO recovery procedure. In Table I selected system-level simulation parameters are shown. In the evaluated scenarios the UE had to measure on the individual radio beams and consolidate them to derive the cell-level measurement that is used in CHO execution monitoring.

TABLE I: Simulation parameters.

| Parameter | Value |
| --- | --- |
| Inter-Site Distance (ISD) | 200 m |
| Deployment | 7 sites, hexagonal cells |
| Tx power | 30 dBm |
| Channel Model | Urban Micro (UMi), compliant with 3GPP Technical Report 38.901 |
| Number of UEs | 420 |
| Carrier frequency | 28 GHz |
| Simulation time | 300 seconds |
| SINR outage limit | -8 dB |
| UE mobility model | Random waypoint |

### A. Conditional Handover in Frequency Range 2

FR2 typically relates to the frequencies above 24 GHz and lower than 50 GHz. Ensuring reliable and robust handover at higher frequency carriers could be more problematic, due to the inherent characteristic of narrow beams. Sudden drops of radio signal level may be encountered in narrow beams (e.g. due to the physical blockage/obstacles between the transmitter and the receiver).

The KPIs which have been derived for assessing the performance are as follows (normalized by the number of UEs and time in unit of minute):

- HO Success (*HOSucc*) logged when the UE successfully completes the handover in the target cell.
- All Mobility Failures (*AllMobilityFail*) logged when the UE experiences a radio link failure in the source cell or fails to successfully complete the handover in the target cell.
- Ping Pong (*PP*) logged when the UE returns to the source cell within 1000 ms since last handover.

First set of results (depicted in Fig. 4 – Fig. 9) compares the performance of basic HO (*BHO*) with CHO for different preparation ($o_{prep}$) and execution ($o_{exec}$) offsets, separately for the UE velocities of 3, 30 and 60 km/h and for the cases when just a single CHO cell or up to four CHO cells are prepared.

As it can be noticed in Fig. 4 and Fig. 7, FR2 UE mobility at 3 km/h does not lead to excessive number of failures. CHO does not bring any clear gain compared to BHO when looking at the results for All Mobility Failures or HO Success. Ping-pongs are also not visible for this UE speed. Eventually, allowing up to 4 CHO candidate cells (Fig. 7) does not lead to increased number of successful HOs in comparison to a single CHO candidate cell (Fig. 4).

The situation changes when analysing the mobility at 30 km/h (Fig. 5 and Fig. 8). The benefits of using CHO become apparent (i.e. increased number of *HOSucc*, reduced *AllMobilityFail*). Early preparation (with 7 dB $o_{prep}$, where $o_{prep}$ denotes how many dBs the source cell was better than candidate) brings further improvement for successful HOs when compared to 3 dB $o_{prep}$. However, it has to be noted such in-advance CHO preparation is inefficient in terms of network resource reservation. Executing CHO late (with 6 dB $o_{exec}$, where $o_{exec}$ denotes how many dBs the candidate was better than source) reduces the PP rate, but simultaneously leads to more mobility failures and decreased amount of successful HOs. Finally, also for 30 km/h there is no gain observed when additional CHO candidate cells are prepared, as shown in Fig. 8. It confirms that for FR2 the main benefit of CHO stems from early preparation of handover, not from preparing multiple cells in parallel. The same has been identified for FR1 in [3].

For 60 km/h (Fig. 6 and Fig. 9) the trends initially observed in Fig. 5 and Fig. 8 become even more evident. The number of mobility failures (*AllMobilityFail*) increases for BHO, but is 3-4 times lower for CHO, depending on the configured $o_{exec}$ and $o_{prep}$. The increased *HOSucc* for early preparation (7 dB $o_{prep}$) is more evident than it was for 30 km/h. Additionally, at 60 km/h the benefit of using 4 CHO candidate cells (Fig. 9) starts to be visible when analysing the *HOSucc* and *AllMobilityFail* and comparing them with a single CHO candidate scenario (Fig. 6).

Overall, CHO in FR2 provides large improvement to mobility KPIs, especially for higher UE velocities considered in this analysis (i.e. 30 km/h and 60 km/h). 7 dB $o_{prep}$ slightly helps in increasing the *HOSucc* ratio, but is not a recommended solution, when its inherent excessive resource reservation is taken into account. This may be especially problematic when multiple CHO candidate cells are prepared. Optimal $o_{exec}$ value is 3 dB. Larger setting, such as 6 dB, reduces the amount of PP, but does not ensure similarly high number of successful HOs. Preparing more than a single CHO candidate cell is not beneficial for slow UEs (i.e. 3 km/h and 30 km/h), minor gains start to be seen for 60 km/h.

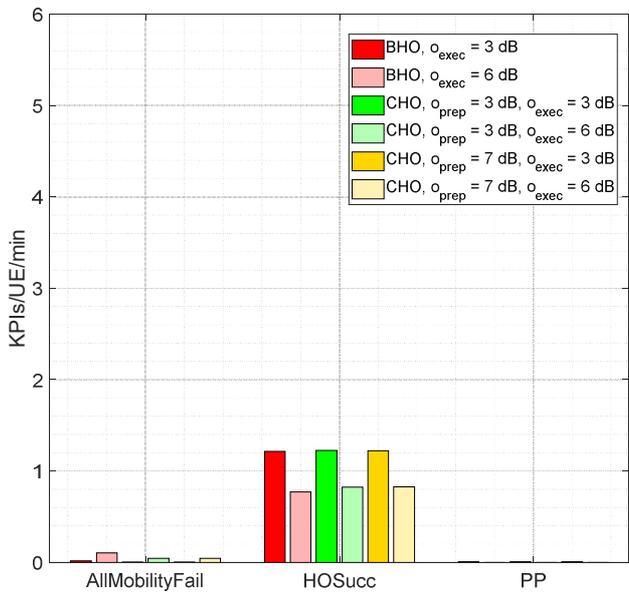

Fig. 4: Mobility KPIs for UE moving at 3 km/h and max. 1 prepared cell

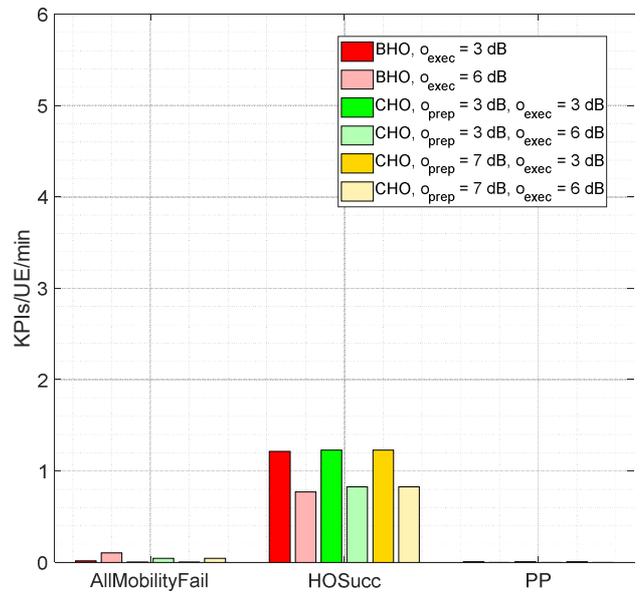

Fig. 7: Mobility KPIs for UE moving at 3 km/h and max. 4 prepared cells

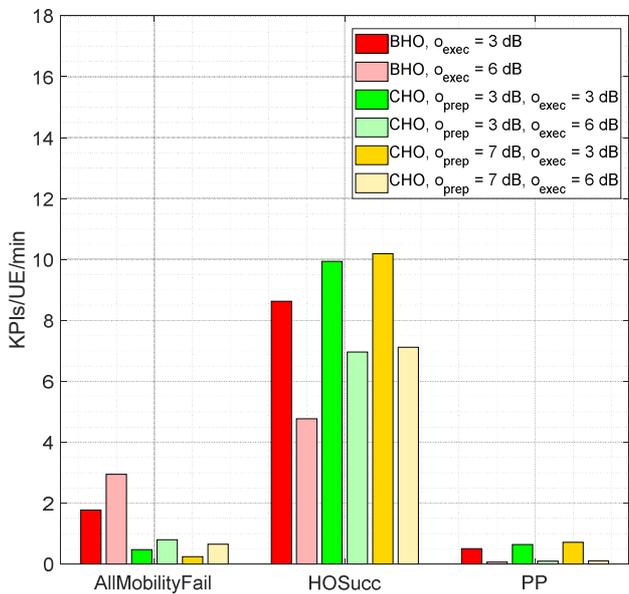

Fig. 5: Mobility KPIs for UE moving at 30 km/h and max. 1 prepared cells

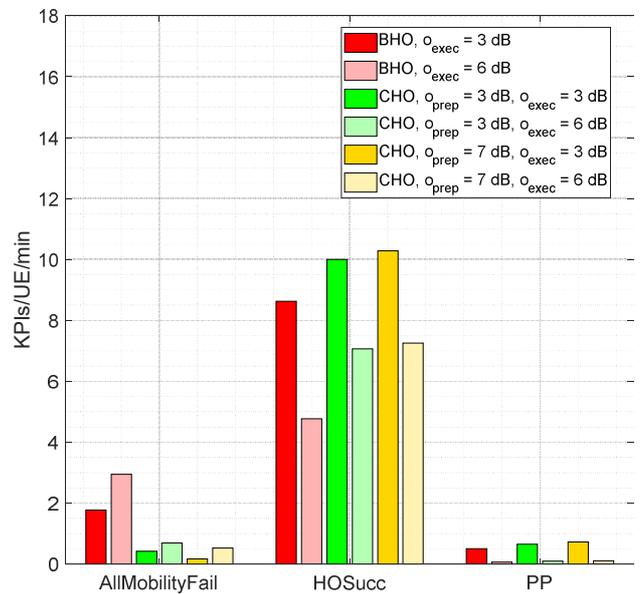

Fig. 8: Mobility KPIs for UE moving at 30 km/h and max. 4 prepared cells

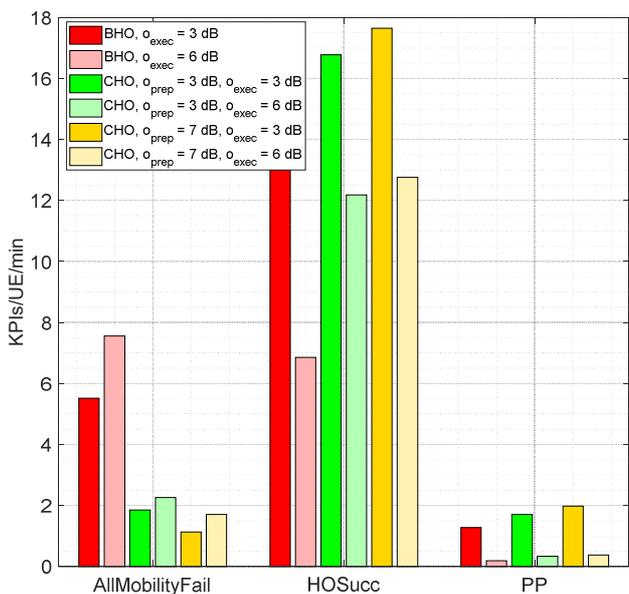

Fig. 6: Mobility KPIs for UE moving at 60 km/h and max. 1 prepared cell

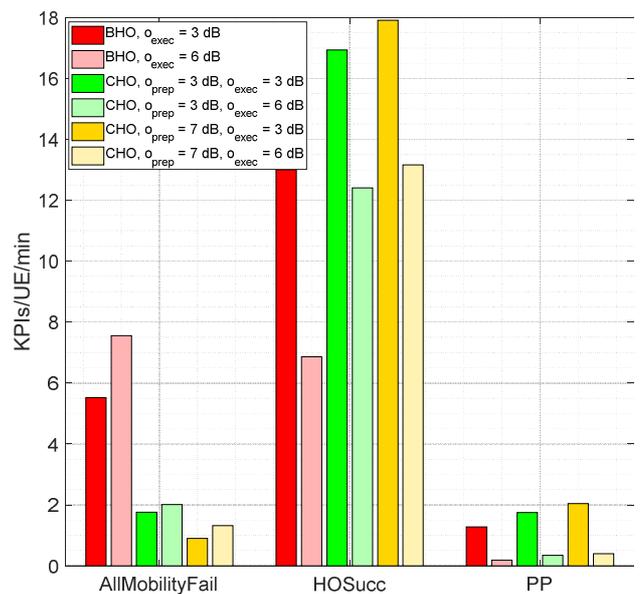

Fig. 9: Mobility KPIs for UE moving at 60 km/h and max. 4 prepared cells

## B. Conditional Handover Recovery Procedure

CHO Recovery has been described in Section II. Here we provide the simulation results showing how this solution performs for different UE velocities (30 km/h and 60 km/h) and various CHO $o_{prep}$ and $o_{exec}$. The results are obtained for different values of the maximum number of CHO prepared cells, i.e. 1, 2, 4 and 8. Vertical axis (*CHO Recovery Rate*) in Fig. 10 and Fig. 11 depicts the percentage of failures resolved using CHO recovery procedure. These numbers have been obtained by dividing the failure cases that are successfully resolved using CHO recovery by the total number of encountered mobility failures. As can be noticed in Fig. 10 the CHO Recovery Rate is higher for cases with 6 dB $o_{exec}$. That is due to the larger overall number of mobility failures and vast majority of them occur when the UE is prepared with multiple CHO candidate cells. Even for lower $o_{prep}$ and $o_{exec}$ values (purple curve in Fig. 10) CHO Recovery helps in resolving up to approximately 25% of failures, what can accelerate the cell access. The number of prepared CHO candidate cells (horizontal axis in Fig. 10 and Fig. 11) impacts the rate of CHO Recovery, especially in the range between 1 and 4, which for the top two curves means the increase from less than 50% for one cell to more than 80 % when up to four cells are prepared (Fig. 10). Preparing more than 4 cells (in our investigated case – 8 CHO candidates) does not bring nearly any benefit, what is consistently proven in all depicted scenarios.

The behavior visible in Fig. 11 is aligned with what has been observed for 30 km/h (Fig. 10). Nevertheless, the CHO Recovery rate for 60 km/h is lower than for 30 km/h. It could be due to the increased number of mobility failures occurring before any CHO preparations are done or as the selected cell in the CHO Recovery process is not one of those which were configured to the UE. Higher UE velocity can quickly make CHO preparations obsolete and non-relevant for CHO Recovery procedure. However, even in the worst case shown in Fig. 11 CHO Recovery rate is above 10% (if at least two CHO candidates are considered), which can be helpful in further improving mobility robustness, largely ensured by CHO already.

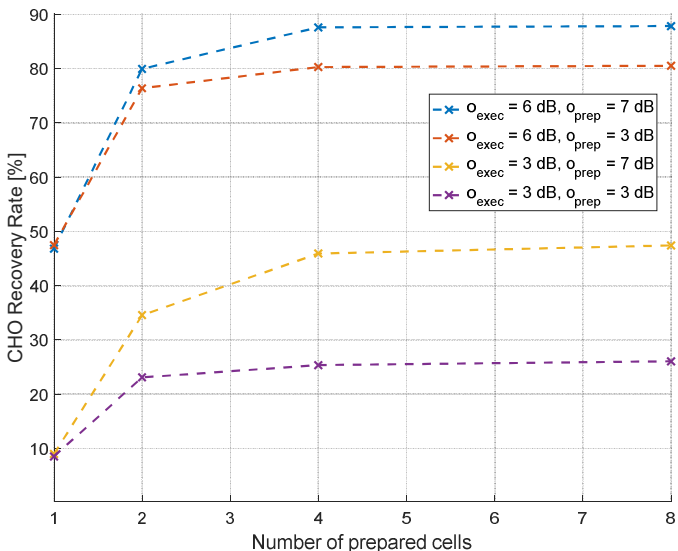

Fig. 10: CHO Recovery rate as a function of the number of prepared CHO cells (30 km/h)

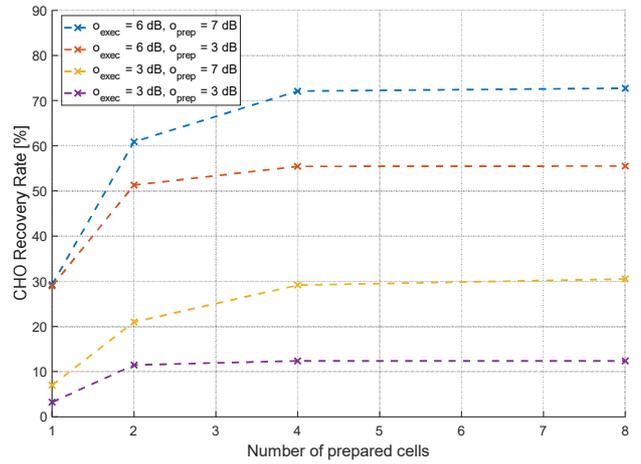

Fig. 11: CHO Recovery rate as a function of the number of prepared CHO cells (60 km/h)

## V. CONCLUSIONS

In this article we have described various aspects of 5G conditional handover, including its principles, applicability to verticals and performance results in FR2. We have presented specific solutions that are designed or may be considered to support CHO, e.g., in NTN or NR-U. We have also investigated mobility KPIs for CHO in FR2. It has been found that CHO clearly reduces the number of failures and increases the number of successful HOs. Such behavior becomes especially evident with the increasing UE's speed. Similar to FR1, also in FR2, the substantial gain of using CHO is visible already when a single candidate cell is prepared. Our analysis has also focused on CHO recovery mechanism, whose usefulness has been evaluated using system-level simulations. It has been depicted CHO Recovery may be applied in up to 90% of failure cases, depending on the scenario and the number of prepared cells. However, for the typical CHO execution and preparation offsets, the gains of CHO Recovery are much lower. CHO attractiveness, both commercial and in research, remains to be high. Undoubtedly its performance will be further improved, and new application areas will be identified. We expect future work to be pursued on combining CHO with other mobility solutions reducing the interruption time, enhancing CHO performance in FR2 and applying CHO in multi-connectivity scenarios. All aimed at eliminating the failures, achieving low interruption period and boosting the achievable data rates.